\documentclass[journal=jacsat,manuscript=article]{achemso}

\usepackage[version=3]{mhchem} 

\author{Joshua Chen}
\email{chenjosh@mit.edu}
\affiliation[1]{Research Laboratory of Electronics, MIT, Cambridge, MA}

\author{Sachin Vaidya}
\affiliation[1]{Research Laboratory of Electronics, MIT, Cambridge, MA}
\alsoaffiliation[2]{Department of Physics, MIT, Cambridge, MA}

\author{Simo Pajovic}
\affiliation[3]{Department of Mechanical Engineering, MIT, Cambridge,  MA}

\author{Seou Choi}
\affiliation[1]{Research Laboratory of Electronics, MIT, Cambridge, MA}

\author{William~Michaels}
\affiliation[1]{Research Laboratory of Electronics, MIT, Cambridge, MA}

\author{Louis Martin-Monier}
\affiliation[4]{Department of Materials Science and Engineering, MIT, Cambridge, MA}

\author{Juejun Hu}
\affiliation[4]{Department of Materials Science and Engineering, MIT, Cambridge, MA}

\author{Carol Cogswell}
\affiliation[5]{Department of Electrical, Computer and Energy Engineering, University of Colorado Boulder, Boulder, CO}

\author{Charles~Roques-Carmes}
\affiliation[1]{Research Laboratory of Electronics, MIT, Cambridge, MA}
\alsoaffiliation[6]{Ginzton Laboratory, Stanford University, Stanford, CA}

\author{Marin Soljačić}
\affiliation[1]{Research Laboratory of Electronics, MIT, Cambridge, MA}
\alsoaffiliation[3]{Department of Physics, MIT, Cambridge, MA}

\title[An \textsf{achemso} demo]
  {Wavefront Engineering for \\ Scintillation-Based Imaging}

\begin{document}

\begin{tocentry}

Some journals require a graphical entry for the Table of Contents.
This should be laid out ``print ready'' so that the sizing of the
text is correct.

Inside the \texttt{tocentry} environment, the font used is Helvetica
8\,pt, as required by \emph{Journal of the American Chemical
Society}.

The surrounding frame is 9\,cm by 3.5\,cm, which is the maximum
permitted for  \emph{Journal of the American Chemical Society}
graphical table of content entries. The box will not resize if the
content is too big: instead it will overflow the edge of the box.

This box and the associated title will always be printed on a
separate page at the end of the document.

\end{tocentry}

\begin{abstract}
Recent research in nanophotonics for scintillation-based imaging has demonstrated promising improvements in scintillator performance. In parallel, advances in nanophotonics have enabled wavefront control through metasurfaces, a capability that has transformed fields such as microscopy by allowing tailored control of optical propagation. This naturally raises the following question, which we address in this perspective: can wavefront-control strategies be leveraged to improve scintillation-based imaging?
To answer this question, we explore nanophotonic- and metasurface-enabled wavefront control in scintillators to mitigate image blurring arising from their intrinsically diffuse light emission. While depth-of-field extension in scintillation faces fundamental limitations absent in microscopy, this approach reveals promising avenues, including stacked scintillators, selective spatial-frequency enhancement, and X-ray energy–dependent imaging. These results clarify the key distinctions in adapting wavefront engineering to scintillation and its potential to enable tailored detection strategies. 
\end{abstract}

\section{I. Introduction}
Scintillation is the process by which certain materials, called scintillators, absorb ionizing radiation such as X-ray or $\gamma$-ray photons and re-emit the deposited energy as visible light detectable by standard photodiodes or camera sensors. This enables indirect measurement of radiation that is otherwise difficult to detect~\cite{weber2002inorganic}. Scintillators are therefore widely used as radiation detectors and are central components in many medical imaging systems, including X-ray radiography, computed tomography (CT), and positron emission tomography (PET). In these imaging applications, scintillator performance metrics such as light yield and absorption efficiency directly impact clinical image quality, radiation dose, and ultimately, the effectiveness of disease diagnosis and treatment.

Recent work has shown that nanophotonic structuring can substantially change how scintillators emit light and, in turn, improve detector performance. By patterning the scintillator or an adjacent layer on the scale of the scintillation wavelength, one can reshape the photonic environment in two main ways: (i) nanostructures can modify the local density of optical states seen by the scintillation transitions, speeding up and spectrally reshaping spontaneous emission~\cite{kurman2020photonic, kurman2024purcell}; and (ii) they can introduce additional outcoupling channels that redirect light toward the detector \cite{roques2022framework}. This ``nanophotonic scintillator'' approach has been explored using plasmonic structures \cite{bignell2013plasmonic,ye2024nanoplasmonic,liu2017plasmonic}, photonic crystals \cite{long2024nonreciprocal, zhang2017enhanced,singh2018enhanced,kurman2020photonic,roques2022framework, martin2025large}, and inverse-designed nanophotonic stacks \cite{shultzman2023enhanced,min2025end}. Together, these studies have demonstrated up to order-of-magnitude increases in light yield~\cite{roques2022framework,martin2025large,kurman2024purcell,ye2024nanoplasmonic,liu2017plasmonic,zhang2017enhanced,shultzman2023enhanced,bignell2013plasmonic}, improved control over angular emission~\cite{roques2022framework,martin2025large,long2024nonreciprocal}, and scalable routes to patterning~\cite{martin2025large,singh2018enhanced,zhang2017enhanced}. Recent analyses estimate that current CT usage may account for roughly 100,000 radiation-induced cancers per year in the U.S., corresponding to up to $\sim$5\% of all new cancers \cite{Cassella_2025,smith2025projected}, underscoring the urgency of improving detector performance. To date, however, this work has largely focused on controlling emission rate, spectrum, and light extraction, leaving scintillation wavefront engineering largely unexplored.

Wavefront engineering refers to shaping how emitted light propagates through the imaging system and is encoded at the detector. More broadly, wavefront engineering encompasses the deliberate modification of the optical wavefront---typically through pupil-plane phase or amplitude control---to tailor the system point spread function (PSF) for a given imaging task. Rather than maximizing image sharpness alone, wavefront engineering redistributes desired optical information in a controlled manner so that it can be more effectively recovered through computation. In scintillation-based imaging, wavefront engineering can extend beyond the scintillator itself to the full imaging pipeline, encompassing the X-ray source, scintillator architecture, detection optics, and reconstruction algorithms. Fig.~\ref{cpmfig5} provides a schematic overview of this broad design space.

{\makeatletter
 \renewcommand{\fnum@figure}{\textbf{Figure~\thefigure}}
 \makeatother
 \begin{figure}[!htbp]
   \centering
   \includegraphics[width=0.75\columnwidth]{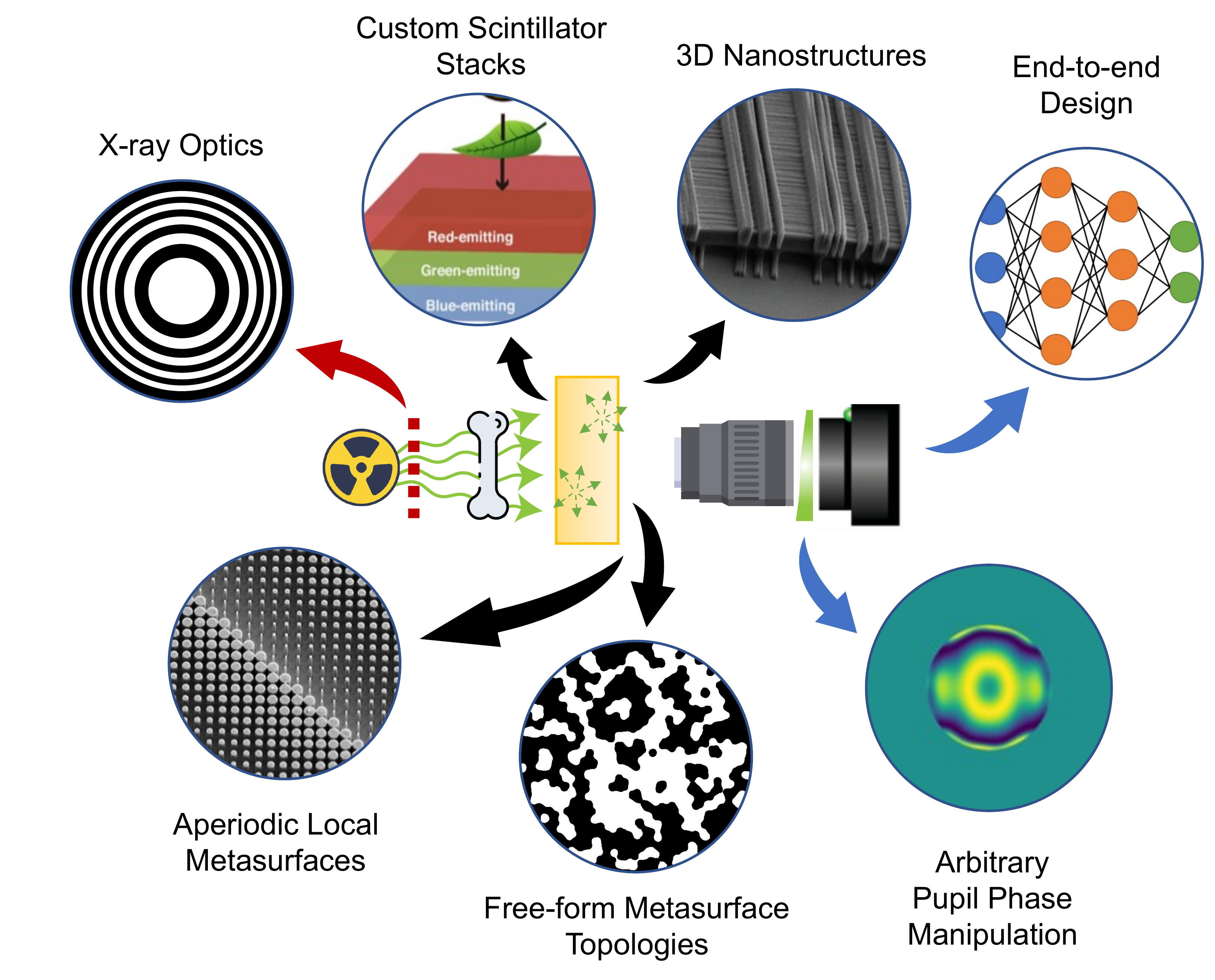}
   {\small
   \caption{\textbf{Wavefront engineering in scintillation-based X-ray imaging.} Schematic highlighting wavefront control in scintillation: tailoring X-ray sources with dedicated optics (here we show a zone plate common for X-ray focusing), designing custom scintillators (e.g., multicolor stacks), and integrating end-to-end computational design. On the detection side, aperiodic, free-form, and 3D nanostructures together with arbitrary pupil-phase control provide further degrees of freedom for shaping scintillation light. Insets for ``Custom Scintillator Stacks'' and ``3D Nanostructures'' are adapted from Refs.~\cite{min2025end}~and~\cite{roques2022toward}.}
   \label{cpmfig5}
   }
 \end{figure}
}

A key enabler of practical wavefront engineering is the emergence of fabrication techniques for metasurfaces and related nanophotonic optical elements that allow compact, lithographically-defined control of optical phase at subwavelength resolution. Aperiodic local metasurfaces~\cite{yu2014flat, chen2016review}, free-form or inverse-designed phase profiles~\cite{liu2018generative, li2022empowering}, and three-dimensional nanophotonic structures~\cite{roques2022toward,melzer20203d,daqiqeh2020nanophotonic} have all been demonstrated at length scales corresponding to the characteristic emission wavelength of scintillators, with many fabrication approaches compatible with wafer-scale processing and large-area patterning. These advances make it feasible to implement sophisticated wavefront-control strategies directly at the detection stage of scintillation systems or in close integration with structured scintillators.

In optical microscopy, wavefront engineering has been widely used to extend imaging capabilities beyond those achievable with conventional, diffraction-limited optics. By shaping the pupil-plane phase, the system PSF can be engineered to encode additional information into the image that is later recovered computationally. Prominent examples include engineered PSFs for three-dimensional single-molecule localization, which encode axial position into the image of a point emitter~\cite{quirin2012optimal,shechtman2014optimal}, as well as adaptive optics, where programmable wavefront corrections compensate aberrations to preserve high resolution deep within biological tissue~\cite{booth2014adaptive}. These approaches illustrate how redistributing optical information---rather than simply sharpening the PSF---can enable new imaging functionalities.


This observation naturally prompts the question of whether analogous wavefront-engineering strategies could address long-standing limitations in scintillation-based X-ray imaging. Scintillation systems face optical-blurring challenges that are reminiscent of depth-of-field constraints in microscopy. To efficiently stop X-rays, scintillators must be millimeters thick, but the visible photons they emit are generated isotropically and can propagate laterally over significant distances before reaching the detector. As scintillator thickness increases, this lateral spread leads to image blur and degraded spatial resolution, creating a fundamental tradeoff between X-ray stopping power and imaging resolution~\cite{shultzman2023enhanced}.

In this perspective, we investigate the use of wavefront engineering for scintillation-based imaging and find that there are fundamental limitations that stem from the depth-distributed nature of scintillation emission. Using wavefront coding as a representative example, we identify when wavefront engineering can provide new functionalities and when fundamental differences between scintillation and conventional microscopy make them ineffective. Guided by these insights, we outline new avenues for wavefront engineering in scintillation, including stacked multicolor scintillators, selective enhancement of specific spatial frequencies, and X-ray-energy-dependent imaging.

\section{II. Wavefront coding: a representative example}

Wavefront coding is a technique that extends the depth of field of an imaging system by deliberately making the PSF invariant over a wide range of depths \cite{dowski1995extended}. Because the PSF hardly changes as the object moves in or out of focus, the recorded image can later be processed with a matching computational operation (e.g., deconvolution) to reconstruct a sharp image over that entire depth range. A seemingly natural extension of wavefront coding to scintillation-based imaging would be to design an imaging system with an extended depth of field that spans the entire scintillator thickness. In principle, such a system would bring emission from all depths into focus simultaneously, allowing their contributions to be combined without blurring from isotropic scintillation light and potentially recovering high-resolution images even in thick scintillators.

{\makeatletter
 \renewcommand{\fnum@figure}{\textbf{Figure~\thefigure}}
 \makeatother
 \begin{figure}[!htbp]
   \centering
   \includegraphics[width=0.5\columnwidth]{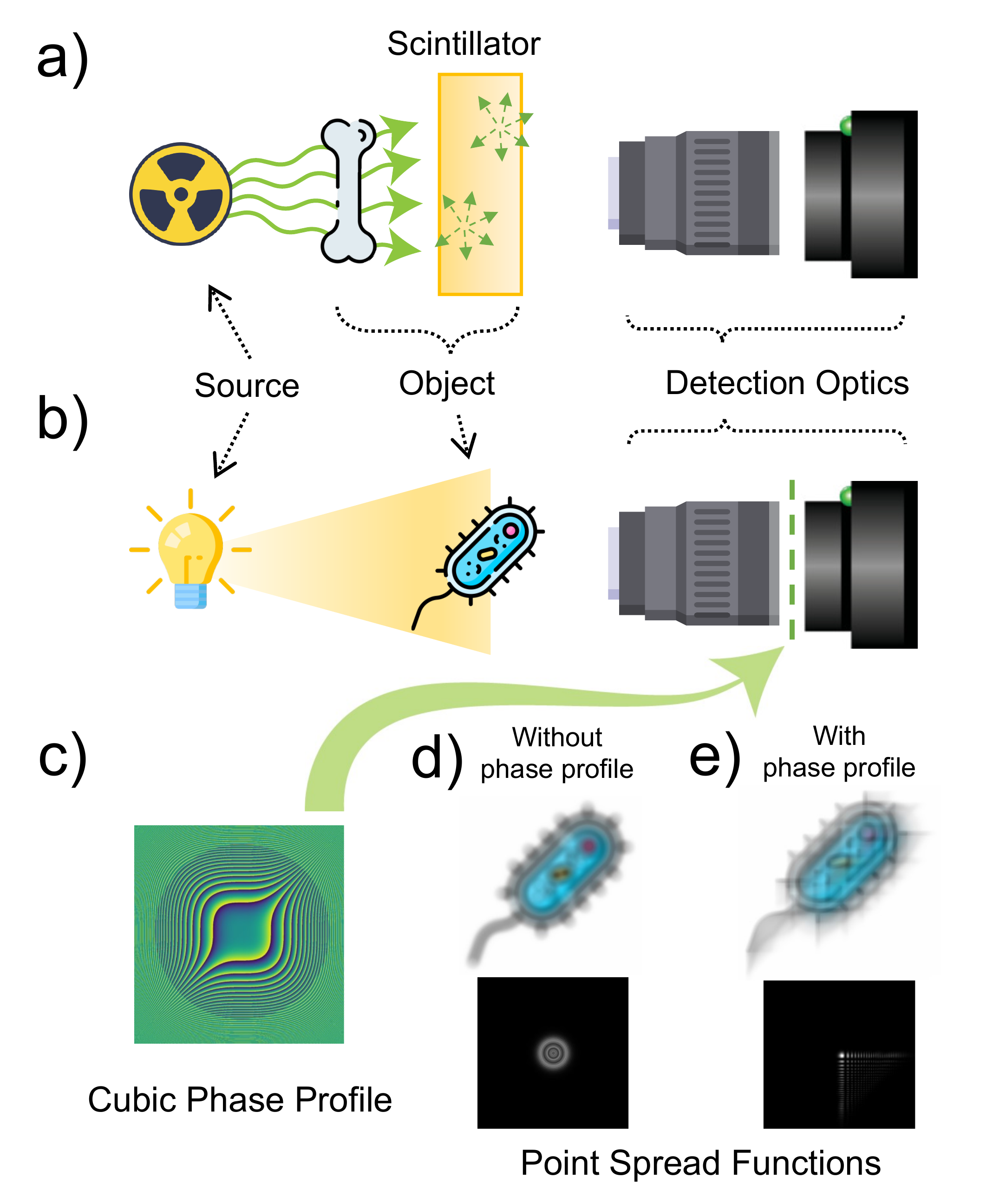}
   {\small
   \caption{\textbf{Similarities between scintillation-based imaging and microscopy.} (a) In scintillators imaged by, for example, free-space optics, light emission from the scintillator is collected and focused to the detector. (b) In microscopy, detection optics similarly image a biological specimen, and wavefront engineering can be used to enhance certain functionalities. (c) A cubic phase profile can be inserted to modify the system point spread function to be depth-invariant. (d) Without a phase profile the PSF results in standard image defocus, but (e) with a cubic phase profile the PSF is altered to be invariant to defocus, allowing extended depth of field (EDOF) to be achieved through computational reconstruction. The ground truth image can be found in (b).}
   \label{cpmfig1}
   }
 \end{figure}
}

Fig.~\ref{cpmfig1} outlines similarities between scintillation-based imaging and microscopy, where a scintillator is coupled to imaging free-space optics. In scintillation-based imaging, the object is encoded in the ``shadow'' of X-rays incident on the scintillator, and emitted according to the depth-dependent characteristics of the scintillator. This is analogous to the imaging of a three-dimensional fluorescent object in emissive microscopy, such as fluorescence microscopy. Furthermore, wavefront coding can be applied to alter the PSF of the system in microscopy, allowing for computational methods to extract desired information. We show the acquired image measurements as well as the associated PSFs with and without a cubic phase mask designed to enable extended depth of field in Fig.~\ref{cpmfig1}d and e. 

Since scintillation is an incoherent emission process, the response of the imaging system is linear in intensity. We model the scintillator as a discrete stack of incoherently emitting planes spanning its thickness, as first shown in Fig.~\ref{cpmfig2}b. Each plane forms an image at the detector according to its associated point spread function, denoted $h_z(\mathbf{r})$ for emission originating at depth $z$, which encodes the depth-dependent optical blur and can be modeled within the framework of Fourier optics~\cite{goodman2005introduction}. The intensity of each plane is set by the fluence of X-rays absorbed at that depth, which can be estimated from the source X-ray spectrum, scintillator X-ray mass attenuation coefficients, and the Beer-Lambert law. This model is validated by a multiphysics Geant4~\cite{agostinelli2003geant4} and Zemax OpticStudio~\cite{ZemaxOpticStudio} ray-tracing pipeline that considers all relevant scintillation physics, described in Supplement~S1. 

While Fig.~\ref{cpmfig1}a depicts a free-space optical readout---common in micro-CT and high resolution laboratory imaging systems---we note that the depth-resolved model introduced above is not specific to this configuration. Many scintillation detectors instead employ direct coupling between the scintillator and the image sensor, where lateral blur arises from near-field scintillator light spreading, internal scattering, and sensor integration rather than from defocus through an imaging lens. In both cases, however, the recorded image is formed as an incoherent sum over emission depth, with each depth contributing a characteristic impulse response $h_z(\mathbf{r})$. The framework developed here therefore applies broadly across detector architectures. Wavefront engineering, whether implemented through free-space meta-optics or via nanophotonic structures patterned directly on or within the scintillator, acts by reshaping these depth-dependent responses and the resulting system PSF.

Fig.~\ref{cpmfig2}a demonstrates the application of wavefront coding with a cubic phase mask to scintillation and compares it with standard detection optics. In the standard case (top), the imaging system has a single focal plane, whereas wavefront coding extends the depth of field across the scintillator (bottom). Fig.~\ref{cpmfig2}b shows the PSFs of select discrete planes along the depth of the scintillator. In the standard case (top row), the PSFs vary strongly in size with depth, while in the wavefront-coded case (bottom row), they exhibit the characteristic cubic PSF shape and remain nearly invariant across the entire thickness. When the emission from all planes is summed, this depth-invariant PSF would, in principle, allow deconvolution to recover the full image. 

{\makeatletter
 \renewcommand{\fnum@figure}{\textbf{Figure~\thefigure}}
 \makeatother
 \begin{figure}[!htbp]
   \centering
   \includegraphics[width=\columnwidth]{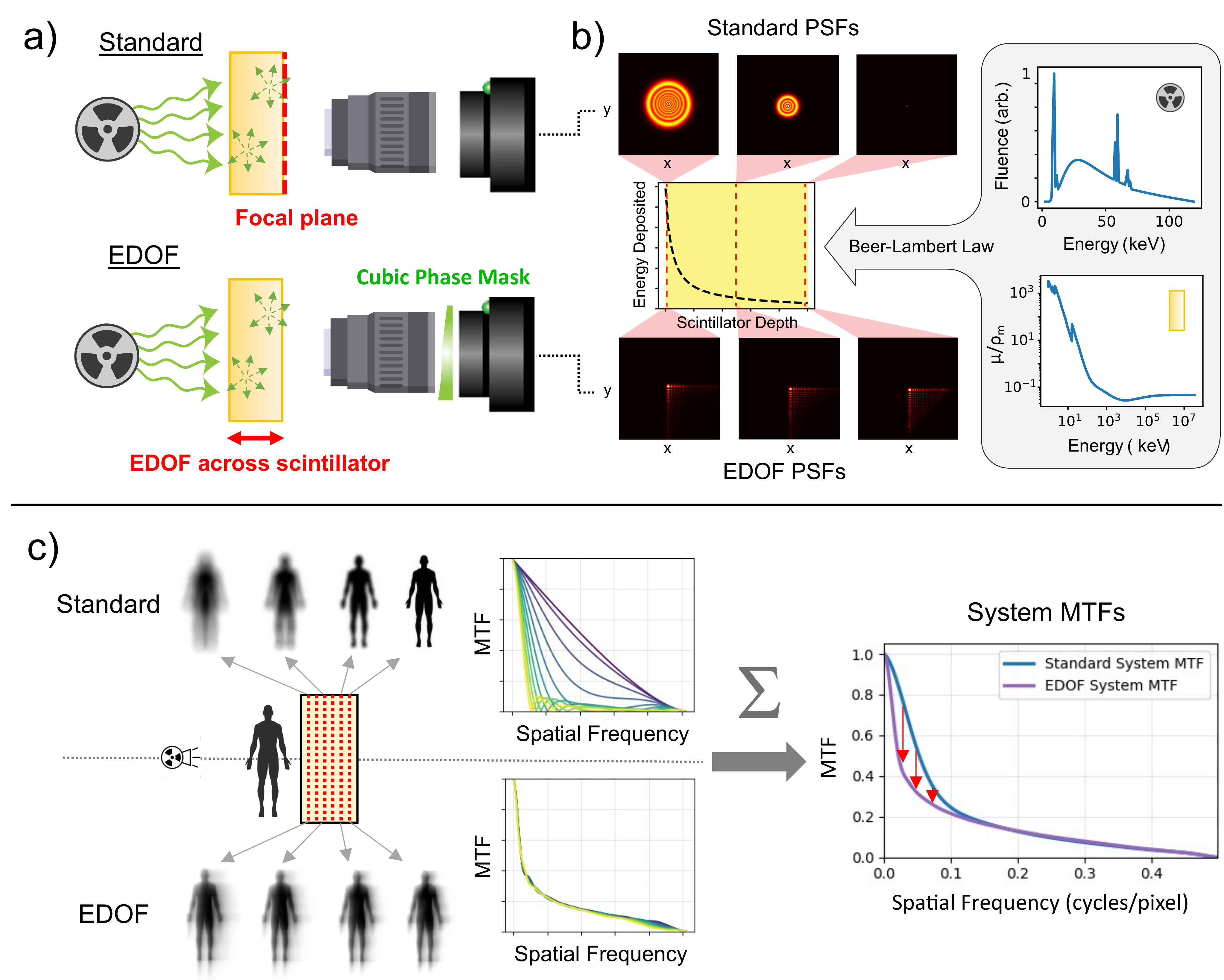}
   {\small
   \caption{\textbf{Extended depth of field (EDOF) for scintillation-based imaging.} (a) Comparison between scintillation using standard detection optics versus scintillation with extended depth of field (EDOF) capabilities. (b) The point spread functions along discrete scintillator planes for standard detection (top) and EDOF detection (bottom). The depth-dependent energy deposition in the scintillator is modeled using the source X-ray spectrum and the energy-dependent scintillator mass attenuation coefficients. (c) Comparison between discrete planar standard and EDOF modulation transfer functions (MTF). Different color lines in the MTF plots correspond to the various discrete planes in the scintillator. Despite the apparent advantage of the discrete planar MTFs in the EDOF case, the system MTF of the standard system lies above the system MTF of the EDOF system at all spatial frequencies.}
   \label{cpmfig2}
   }
 \end{figure}
}

\section{III. Limits of image reconstruction in scintillation-based imaging}

A key distinction between scintillation and conventional imaging modalities like microscopy is the integration of depth information at the detector, which fundamentally alters how wavefront coding can be applied in scintillation-based imaging.
This difference arises once we consider the full imaging workflow, which includes a reconstruction step (typically, deconvolution) to generate a final image. Fig.~\ref{cpmfig2}c illustrates how wavefront coding can shape the modulation transfer function (MTF), which quantifies how an imaging system transfers object contrast to the image as a function of spatial frequency. In the standard imaging system (top plot), the image blur varies with emission depth in the scintillator: the in-focus plane has a high MTF, while out-of-focus planes show a rapid loss of contrast. In the extended depth of field (EDOF) system (bottom plot), the object image remains nearly unchanged for all emission planes because the PSFs are depth-invariant. The corresponding MTFs are likewise consistent across planes and, importantly, exhibit no zeros in spatial frequency, a typical requirement for stable image reconstruction.

However, the picture is different once we include the reconstruction process in the comparison. The key distinction lies in what PSF is used in reconstruction. In microscopy, computational post-processing is performed with a single in-focus PSF, which can be directly measured. For example, in Wiener deconvolution~\cite{kakutani1950nobert}, the PSF is used in its Fourier-space representation as $H_0$:
\begin{equation}
\hat{O}(\mathbf{k}) =
\frac{H_0^*(\mathbf{k})}{\lvert H_0(\mathbf{k}) \rvert^2 + \gamma}\, I(\mathbf{k})
\label{eq:wiener}
\end{equation}
where $\hat{O}(\mathbf{k})$ is the reconstructed image in the spatial-frequency domain, $\mathbf{k}$ denotes the spatial frequency coordinate, $\gamma$ is a regularization parameter related to the noise-to-signal power ratio, and $I(\mathbf{k})$ is the measured image in the Fourier domain. The form of this Wiener deconvolution, along with the underlying assumptions, is summarized in Supplement~S2.

In scintillation, however, the measured image is an incoherent sum of contributions from all depths, so linearity of convolution leads to a single effective system PSF (Fig.~\ref{cpmfig2}b) given by a fluence-weighted sum of the depth-dependent PSFs: 
\begin{equation}
h_{\mathrm{sys}}(\mathbf{r}) = \sum_z w_z\, h_z(\mathbf{r})
\label{eq:hsys}
\end{equation}
where $h_z(\mathbf{r})$ is the PSF for emission originating at depth $z$, and $w_z$ is the corresponding fluence weighting. Since $I(\mathbf{k})$ is approximated by $O(\mathbf{k})\, H_{\mathrm{sys}}(\mathbf{k})$ in scintillation, deconvolution must therefore be performed with the Fourier-space representation $H_{\mathrm{sys}}(\mathbf{k})$, rather than the single-plane PSF $H_{\mathrm{0}}(\mathbf{k})$. Put simply, depth in microscopy carries new information whereas depth in scintillation carries only redundancy and blur, since all scintillator depths image the same transverse object. 

In practice, traditional deconvolution with the system PSF given by Eq.~\eqref{eq:hsys} yields little to no improvement in resolution: the system's MTF is below that of the standard (non-wavefront-coded) system at all spatial frequencies, shown in Fig.~\ref{cpmfig2}c. While widefield fluorescence microscopy also involves incoherent emission from multiple depths, wavefront coding is typically applied in regimes where depth information is either sparse, recoverable, or explicitly reconstructed. In scintillation-based imaging, depth is irreversibly integrated prior to detection, fundamentally altering the role of wavefront engineering. This fundamental difference constrains how wavefront coding can be used for resolution enhancement in scintillation, but it also points to new potential functionalities for wavefront engineering in scintillator design.

\section{IV. Wavefront engineering for scintillation-based imaging}

\subsection{Extended depth of field (EDOF) in scintillation-based imaging}

In this section, we highlight how one can adapt wavefront coding to applications in scintillation-based imaging, leading to new performance metrics and use cases with distinct functionalities. In all of the following examples, we use Wiener deconvolution~\cite{wiener1964extrapolation} for post-processing, since it depends only on the system PSF and noise statistics.{\renewcommand{\thefootnote}{\fnsymbol{footnote}}\setcounter{footnote}{1}\footnote{Other algorithms that incorporate stronger priors may perform better for specific tasks, but we do not consider them here to simplify the comparison.}\setcounter{footnote}{0}}
Throughout, we use the system MTF as the primary performance metric for wavefront-engineered scintillation because it directly quantifies the robustness of each spatial frequency to noise in the system. Since the phase-only pupil modifications considered here do not change the total detected photon flux or the underlying detector noise sources, we assume a design-independent noise power spectrum (NPS), under which the system MTF serves as a straightforward proxy for the fundamental performance metric of detective quantum efficiency (DQE)~\cite{siewerdsen1998signal}.

The traditional use of wavefront coding for depth-of-field extension is most useful when none of the relevant emitting scintillator planes are sharply imaged by the system. If the focus of the imaging system lies inside the scintillator, then some emission planes are already well focused, so wavefront coding only smears information that is already sharp. On the other hand, if the focus lies outside the scintillator, then no emission plane is well focused, and wavefront coding can preserve high-frequency information that would otherwise be lost. Figure~\ref{cpmfig3}a illustrates this point. The shaded yellow band on the left marks the scintillator thickness, and the dashed lines labeled (1)-–(3) indicate three possible focus planes of the objective: in the center of the scintillator, at its edge, and outside it. The three panels on the right show, for each focus position, the system MTF of a standard objective (dotted black) and of the same objective with a cubic phase mask (solid blue). When the focus lies inside the scintillator (cases (1) and (2)), the cubic-mask MTF stays below the standard objective MTF at all spatial frequencies, so wavefront coding only degrades the information that is already in focus. When the focus lies outside the scintillator in case (3), no emission plane is sharply imaged, so the unmasked system strongly suppresses high spatial frequencies in $H_{\mathrm{sys}}$. In this regime, the cubic phase mask redistributes defocus blur more evenly across depth, leading to a larger $\lvert H_{\mathrm{sys}}(\mathbf{k}) \rvert$ at high spatial frequencies than for the standard objective. Thus, extended depth of field through wavefront coding is only beneficial when the defocused information does not already appear sharply at some in-focus plane of the system---a guideline that applies to all applications of wavefront coding. 

{\makeatletter
 \renewcommand{\fnum@figure}{\textbf{Figure~\thefigure}}
 \makeatother
 \begin{figure}[!htbp]
   \centering
   \includegraphics[width=\columnwidth]{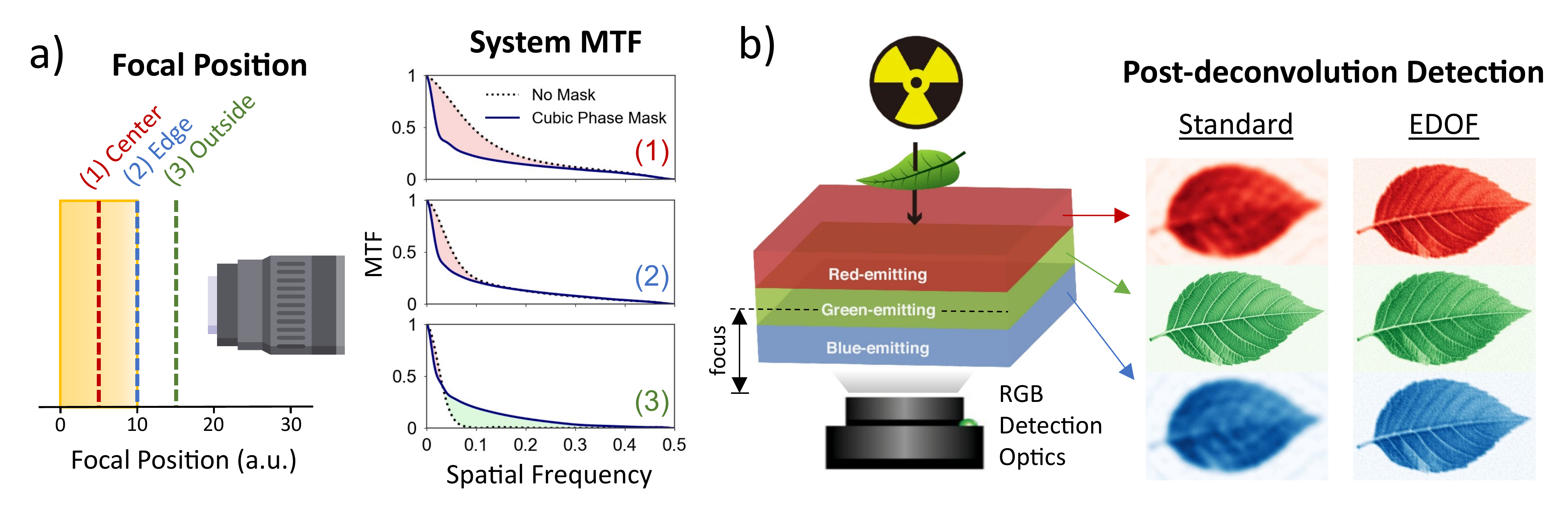}
   {\small
   \caption{\textbf{Use cases for EDOF in scintillation-based imaging.} (a) When the desired focal plane lies outside the scintillator volume, extended depth of field becomes beneficial, as seen in the system MTFs comparing the standard and cubic phase mask designs. (b) The same principle extends to stacked multicolor scintillator systems, where, without EDOF, the scintillator layers away from the chosen focal plane are severely out of focus.}
   \label{cpmfig3}
   }
 \end{figure}
}

This observation has important implications for stacked scintillator systems. Such stacks have been explored for energy-resolving detection, where scintillators with different emission characteristics are combined to recover information about X-ray attenuation \cite{maier2017dual,ran2022multispectral,he2024multi,maurino2016theoretical}, enabling energy resolution and material decomposition \cite{alvarez1976energy}. As with single-layer scintillators, stacked designs still face a tradeoff between thickness and spatial resolution. This tradeoff can be even more severe when light generated in one scintillator must traverse another scintillator layer before reaching the detector.

For example, Ref.~\cite{min2025end} recently proposed a multicolor scintillator stack coupled to a red-green-blue (RGB) detector, consisting of three layers emitting approximately in the red, green, and blue spectral bands. This work offers a concrete example of a stacked scintillator architecture with spectrally separated readout. Fig.~\ref{cpmfig4}b shows a proof-of-concept implementation of such a multicolor stack and the corresponding post-deconvolution images for both a standard imaging system and an EDOF system using a cubic phase profile. As seen in the leaf images, the red- and blue-emitting layers suffer from severe loss of spatial resolution without an EDOF design.

Having identified use cases where traditional wavefront coding benefits scintillation and adopted the system MTF as our figure of merit, we now ask what pupil phase profile might be optimal for scintillation.

\subsection{Inverse design of wavefront encoders for scintillation-based imaging}

Having established the system MTF as our figure of merit, we can use inverse design~\cite{molesky2018inverse} methods to automatically discover a pupil phase profile for enhanced scintillation-based encoding by maximizing a cost function of the system MTF. Rather than restricting ourselves to a specific mask such as a cubic phase profile, we turn to gradient-based inverse design enabled by our auto-differentiable simulation framework introduced in Fig.~\ref{cpmfig2}b. For optimization, we parameterize the pupil phase using the standard Zernike basis \cite{von1934beugungstheorie}, a set of orthogonal polynomials widely used to represent optical aberrations. Zernike modes provide an efficient, non-redundant expansion of pupil phase variations and help reduce parameter coupling during optimization. Using the Zernike basis shown in Fig.~\ref{cpmfig4}a, we optimize the Zernike coefficients up to 15\textsuperscript{th} order (136 basis functions in total) to maximize the system MTF for scintillation. We define the optimization objective as the integrated system MTF over spatial frequencies,
\[
\mathcal{L}_{\mathrm{MTF}} = \int_{0}^{k_{\max}} \mathrm{MTF}_{\mathrm{sys}}(k)\, \mathrm{d}k,
\]
where $\mathrm{MTF}_{\mathrm{sys}}(k)$ is the system MTF and $k_{\max}$ denotes the maximum spatial frequency represented in the discretized Fourier domain. Performing optimization reveals two key insights: 

\begin{enumerate}
  \item \emph{Maximizing the integrated system MTF over all spatial frequencies always shifts the plane of focus to the average intensity depth in the scintillator.}
\end{enumerate}

Among all 136 Zernike basis functions, only the basis function responsible for defocus ($Z^0_2$) significantly affects the integrated MTF, acting effectively as an additional lens term that adjusts the focal length of the system. This means the optimization does not introduce higher-order phase structure, but instead selects a ``plane of best focus'' inside the scintillator. We find empirically that this optimized plane coincides with the average emission depth, given by

\begin{equation}
x_{\mathrm{optimal}}
= \frac{\displaystyle\int_{0}^{d} x\,s(x)\,dx}
       {\displaystyle\int_{0}^{d} s(x)\,dx}
\label{eq:x_optimal}
\end{equation}

\noindent where $s(x)$ is the scintillation energy weighting as a function of depth and $d$ is the scintillator thickness (for example, $s(x)=\frac{1}{m}e^{-\frac{x}{m}}$ for a monochromatic X-ray energy with mean attenuation length $m$). A natural interpretation of this result is that focusing near the average emission depth minimizes the cumulative blur in the final image by balancing the contributions from all emitting planes. This ``optimal plane'' has important implications for wavefront engineering with metasurfaces or other aperiodic nanophotonic structures fabricated directly on scintillators.

{\makeatletter
 \renewcommand{\fnum@figure}{\textbf{Figure~\thefigure}}
 \makeatother
 \begin{figure}[!htbp]
   \centering
   \includegraphics[width=\columnwidth]{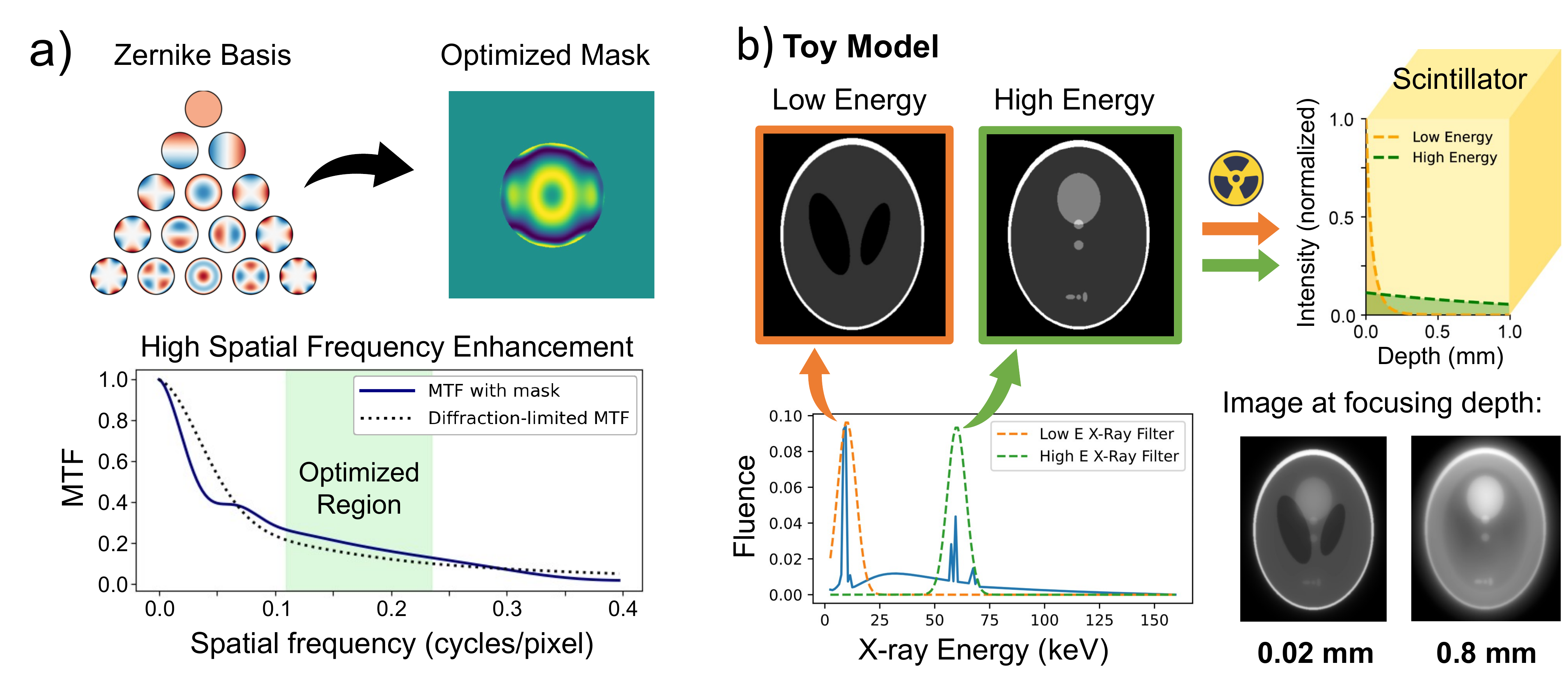}
   {\small
   \caption{(a) Inverse design using the Zernike basis can optimize for certain spatial frequencies above the in-focus optimal MTF. (b) A toy model demonstrates energy-dependent scintillation imaging, where some features of the Shepp-Logan phantom are contained by high energy X-rays and some features are contained by low energy X-rays. The X-ray energy deposited for each portion of the Shepp-Logan phantom is plotted on the yellow rectangular prism representing the scintillator. The detected image thus varies according to the plane of focus in the scintillator.}
   \label{cpmfig4}
   }
 \end{figure}
}

\begin{enumerate}
  \setcounter{enumi}{1} 
  \item \emph{Certain spatial frequencies can be enhanced through inverse design.}
\end{enumerate}

In conventional wavefront coding, no phase mask can increase the system MTF at any spatial frequency beyond the in-focus, diffraction-limited MTF because the two-dimensional optical transfer function (whose magnitude is the MTF) is the autocorrelation of the pupil function within the paraxial framework \cite{goodman2005introduction}. Any additional phase in the pupil can only introduce destructive interference and reduce the MTF at various spatial frequencies: as derived in Supplement~S3 there exists a bound
\begin{equation*}
\mathrm{MTF}(\mathbf{k}) \;\le\; \mathrm{MTF}_{\mathrm{DL}}(\mathbf{k})
\qquad \forall\,\mathbf{k},
\end{equation*}
where $\mathbf{k}$ denotes the transverse spatial frequency vector in the image plane, $\mathrm{MTF}(\mathbf{k})$ denotes the modulation transfer function for an arbitrary pupil phase, and $\mathrm{MTF}_{\mathrm{DL}}(\mathbf{k})$ is the diffraction-limited MTF corresponding to a uniformly phased pupil, for which the magnitude of the pupil autocorrelation is maximized.

In scintillation systems, however, this constraint can be relaxed because the image formation process is intrinsically depth-dependent. As shown in Fig.~\ref{cpmfig4}a, when we choose an objective function that emphasizes a specific spatial-frequency band, inverse design yields phase profiles that boost the system MTF above the in-focus MTF of the standard system within that band, at the expense of spatial frequencies outside the optimized range. This targeted enhancement suggests that scintillation systems can be tailored to preferentially transmit particular spatial scales---corresponding to specific feature or object sizes such as anatomical structures, lesions, or growths---with higher fidelity than would otherwise be possible.

Building on these insights from inverse design, we next examine how the optimal plane shifts when the depth-weighting $s(x)$ changes with X-ray energy and how this can be exploited for energy-dependent imaging.

\subsection{X-ray energy encoding with wavefront engineering}

Up to this point, we have treated the scintillator response as if it were independent of distinct X-ray energies. However, identifying distinct X-ray energies can provide substantial additional information, enabling material discrimination and quantitative imaging~\cite{he2024multi,ran2022multispectral,alvarez1976energy}. In our framework, X-ray photons of different energies are associated with different scintillator mass attenuation coefficients, so they deposit energy at different depths and thus contribute different depth weightings $s(x)$. Through Eq.~\ref{eq:x_optimal}, this immediately implies that each X-ray energy (or energy band) has its own optimal plane of focus inside the scintillator, given by its own average emission depth.

To illustrate this, we consider a simplified toy model in which different spatial features of an object are carried by different X-ray energy bands (whereas in reality, low- and high-energy photons will be correlated through the material mass attenuation coefficients). Fig.~\ref{cpmfig4}b shows a segmented Shepp-Logan phantom~\cite{shepp1974fourier} in which small features are associated with higher-energy X-rays and large features with lower-energy X-rays. Because the energy-dependent attenuation coefficients differ across the segments, each band produces a distinct depth profile $s(x)$, and therefore a different optimal focus plane according to Eq.~\eqref{eq:x_optimal}. Guided by this equation, we select two imaging depths in a 1~mm-thick scintillator: one near the front facet at 0.02~mm and one near the back facet at 0.80~mm. As seen in the simulated images, changing the focus between these depths selectively sharpens different sets of features, corresponding to different energy bands and thus different material information.

This example shows that wavefront engineering in scintillation exposes a new handle for energy-dependent imaging: the choice of focal plane implicitly selects an ``average'' X-ray energy and the associated material information encoded at that depth. The depth–energy structure of scintillation then becomes a tunable channel, where focusing at different planes emphasizes different energy bands and, in turn, different features or materials in the object.

\section{V. Outlook}

Our results show that wavefront engineering in scintillation must be approached differently than in microscopy. As shown, direct application of wavefront-coding strategies fails because scintillation images are formed as incoherent sums over emission depth. When viewed at the system level, however, wavefront engineering still opens a rich and structured design space for scintillator-based imaging. By working with the system PSF and MTF, we identified when extended depth of field is actually beneficial, how depth-dependent blurring in stacked scintillators can be mitigated, how inverse design reveals an ``optimal'' focus plane set by the depth distribution of scintillation, and how specific spatial frequencies can be selectively emphasized. We also showed that these ideas naturally extend to X-ray-energy-dependent imaging, where the optimal wavefront pattern is tied to the average emission depth of a given energy band.

We highlight a broad landscape of directions for advancing scintillation-based imaging with influence from wavefront engineering. Across the imaging pipeline, new degrees of freedom are becoming accessible that make it possible to shape, encode, and interpret scintillation light in ways that have yet to be explored. On the detector side, recent demonstrations of nanophotonic structuring directly within scintillators---including volumetric and three-dimensional architectures---establish the feasibility of embedding complex functionality at the point of emission~\cite{jurgensen2025volumetrically}. More broadly, continued progress in nanofabrication, including nanoimprint lithography~\cite{chou1996nanoimprint}, colloidal and block-copolymer self-assembly~\cite{zhang2010colloidal, mai2012self}, and two-photon lithography~\cite{wu1992two}, is expanding the set of wavefront- and emission-shaping structures that can be realized in or on scintillators. A range of material platforms have already been used for optical phase and scattering control in this context, including dielectric platforms such as SiN, TiO$_2$, and SiO$_2$ patterned as layers on scintillators, direct structuring of common scintillator materials (e.g., garnets and other oxides), and structured templates that can be infused with liquid scintillators to realize composite architectures. In parallel, emerging work on stacked and other architected scintillator systems demonstrates how depth, spectrum, and emission pathways can be co-designed at the material level, providing a complementary axis for encoding information prior to detection~\cite{min2025end}. Beyond the scintillator itself, pupil-phase manipulation and end-to-end optical-computational co-design offer additional levers for task-specific encoding and reconstruction~\cite{chen2025phase,lin2022end,roques2025metaoptic}. Similar principles could also be extended upstream, where engineered X-ray illumination or source-side structuring might be co-designed with scintillator architectures and detection optics to better align scintillation light for specific imaging tasks. Taken together, these directions point toward a shift from treating scintillation as a fixed light source to viewing it as an integral, designable component of the imaging system, with wavefront engineering providing a unifying framework for understanding performance limits and guiding future detector architectures.

\section{Acknowledgments}
The authors would like to thank Dr. Tom Vettenburg, Dr. Steven E. Kooi, and Chris Hogan for stimulating discussions. This work was supported by the U.S. Army Research Office through the Institute for Soldier Nanotechnologies at MIT under Collaborative Agreement Number W911NF-23-2-0121. J.C. and W.M. acknowledge support from the NSF GRFP under Grant Number 2141064. S.P. acknowledges the MathWorks Engineering Fellowship. S.C. acknowledges support from the Korea Foundation for Advanced Studies Overseas PhD Scholarship. C.~R.-C. is supported by a Stanford Science Fellowship. ChatGPT (OpenAI) and Gemini (Google) were used for clarity and language edits; all scientific content and conclusions are those of the authors.


\bibliography{achemso}

\end{document}